\documentclass[sigconf,nonacm]{acmart}

\setcopyright{none}
\usepackage{booktabs}
\usepackage{amsmath}
\usepackage{graphicx}
\usepackage{multirow}
\usepackage{enumitem}

\newcommand{\pace}{PACERec}

\newcommand{\R}{\mathbb{R}}

\title[PACERec]{PACERec: Progressive Adaptation of LLM Item Embeddings with Collaborative Evidence for Recommendation}

\author{Junzhao Huang}
\authornote{These authors contributed equally.}
\email{huangjunzhao@mails.neu.edu.cn}
\author{Xi Wu}
\authornotemark[1]
\email{wux2@mails.neu.edu.cn}
\author{Zihan Wang}
\author{Zijing Wang}
\author{Ming Wang}
\author{Kang Liu}
\author{Xiaocui Yang}
\author{Shi Feng}
\authornote{Corresponding author.}
\email{fengshi@cse.neu.edu.cn}
\author{Daling Wang}
\author{Yifei Zhang}
\affiliation{
  \department{School of Computer Science and Engineering}
  \institution{Northeastern University}
  \city{Shenyang}
  \country{China}
}
\author{Ting Yang}
\author{Xurui Li}
\affiliation{
  \institution{Ant Group}
  \country{China}
}

\makeatletter
\renewcommand{\@mkauthors@iii}{%
  \gdef\@currentauthors{}%
  \gdef\@currentaffiliation{}%
  \global\setbox\mktitle@bx=\vbox{%
    \unvbox\mktitle@bx\par\medskip
    \centering
    {\large\sffamily
      Junzhao Huang\textsuperscript{1,*}\quad
      Xi Wu\textsuperscript{1,*}\quad
      Zihan Wang\textsuperscript{1}\quad
      Zijing Wang\textsuperscript{1}\\[3pt]
      Ming Wang\textsuperscript{1}\quad
      Kang Liu\textsuperscript{1}\quad
      Xiaocui Yang\textsuperscript{1}\quad
      Shi Feng\textsuperscript{1,\textdagger}\\[3pt]
      Daling Wang\textsuperscript{1}\quad
      Yifei Zhang\textsuperscript{1}\quad
      Ting Yang\textsuperscript{2}\quad
      Xurui Li\textsuperscript{2}\par}
    \medskip
    {\normalsize
      \textsuperscript{1}School of Computer Science and Engineering,\\
      Northeastern University, Shenyang, China\\[2pt]
      \textsuperscript{2}Ant Group, China\par}
    \smallskip
    {\small
      \href{mailto:huangjunzhao@mails.neu.edu.cn}{\texttt{huangjunzhao@mails.neu.edu.cn}}
      \quad
      \href{mailto:wux2@mails.neu.edu.cn}{\texttt{wux2@mails.neu.edu.cn}}
      \quad
      \href{mailto:fengshi@cse.neu.edu.cn}{\texttt{fengshi@cse.neu.edu.cn}}\par}
    \bigskip
  }%
}
\makeatother

\begin{document}

\begin{abstract}
Large language model (LLM) embeddings have been increasingly adopted in sequential recommendation to encode static item attributes and alleviate the limitations of item ID embeddings under sparse interactions.
However, these embeddings typically capture only static content similarity while neglecting sequential transitions, and since their adaptation quality relies on sparse per-item updates or global geometric calibration, existing methods still struggle to explicitly transfer preceding interaction contexts from frequent to long-tail items under unevenly distributed behavioral evidence, limiting recommendation effectiveness across the catalog.
To address this challenge, we propose \textbf{PACERec}, an ID-free progressive adaptation framework that incorporates preceding interaction contexts into semantic representation adaptation by learning a shared velocity field from items with sufficient interactions via frequency-weighted flow matching.
In this way, PACERec progressively guides static LLM embeddings toward collaborative behavioral targets while preserving original item semantics, overcoming sparse per-item updates to effectively generalize sequential transition patterns to long-tail items for both user modeling and candidate scoring.
Extensive experiments on three real-world datasets across three sequential backbones demonstrate the superiority of our method over competitive baselines in both overall and long-tail recommendation.
\end{abstract}

\ccsdesc[500]{Information systems~Recommender systems}

\keywords{Item Representation, Flow Matching, Large Language Models, Long-Tail Sequential Recommendation}

\maketitle

\section{Introduction}

\begin{figure}[t]
    \centering
    \includegraphics[width=\columnwidth]{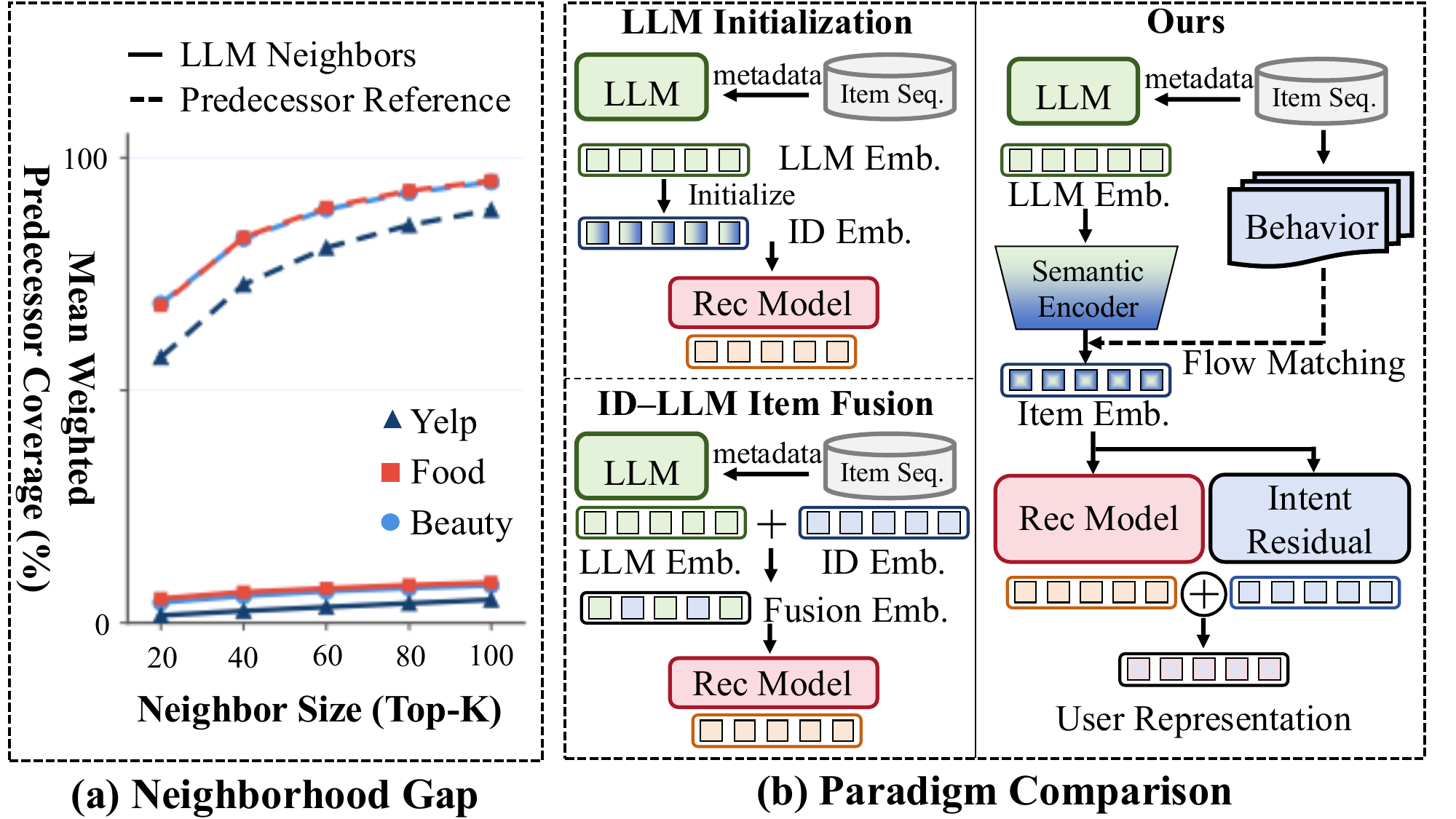}
    \caption{(a) Mean weighted training predecessor coverage on Yelp, Food, and Beauty.
Solid curves use the $K$ nearest neighbors in the frozen LLM embedding space.
Dashed curves use each item's $K$ most frequent training predecessors.
(b) LLM initialization and ID--LLM fusion versus PACERec's
behavior-guided semantic item adaptation and intent-residual user modeling.}
    \Description{Two panels: (a) line chart of mean weighted predecessor coverage on Yelp, Food, and Beauty; (b) architecture comparison between LLM initialization, ID-LLM fusion, and PACERec.}
    \label{fig:motivation}
\end{figure}

Incorporating large language models (LLMs) into sequential recommendation has become a prominent paradigm across modern Web platforms~\cite{liu2024llmesr,hu2025alphafuse,wei2026faerec}, where frozen LLM item embeddings are extracted from static item attributes to compensate for sparse collaborative signals.
Compared with conventional models that learn item-specific ID embeddings solely from interaction data~\cite{hidasi2016session,tang2018caser,sasrec,bert4rec}, leveraging these content-derived embeddings alleviates the limited collaborative supervision on tail items under data sparsity~\cite{NIPS2017_dbd22ba3}, which would otherwise degrade the item representations needed to capture sequential dependencies and rank candidates effectively.
However, because frozen LLM item embeddings are derived solely from static item content, they capture only stable content similarity, which conflicts with the progressive preference transitions in user interactions~\cite{mcauley2015inferring,he2016fusing}.
For example, after purchasing a digital camera, a user may subsequently seek a compatible lens or memory card rather than another camera with a similar description, illustrating how semantic similarity alone fails to capture evolving user needs.
To quantify this discrepancy, we analyze 2,560 frequency-stratified items from each of the Yelp, Food, and Beauty training sets and measure how well an item's size-$K$ semantic neighborhood covers its training predecessors (items interacted with shortly before this item).
Figure~\ref{fig:motivation}(a) shows that at $K=20$, the nearest semantic neighbors from frozen LLM item embeddings (solid lines) cover only 1.59\%--5.22\% of training predecessors, versus 57.09\%--68.65\% for the predecessor reference formed by the 20 most frequent training predecessors (dashed lines).
Even at $K=100$, semantic neighbors reach only 4.97\%--8.66\% coverage compared with 88.80\%--95.08\% for the predecessor reference.

This large discrepancy indicates that relying solely on static LLM item embeddings struggles to capture collaborative transition patterns across user interactions, limiting recommendation performance.
Recent studies have begun exploring ways to integrate collaborative signals with LLM item embeddings so that sequential recommenders can benefit from both semantic and behavioral information.
Existing approaches generally fall into two categories depending on whether they retain item-specific ID embeddings.
As illustrated in Figure~\ref{fig:motivation}(b), (i) one line of work retains item-specific ID embeddings via LLM initialization or ID--LLM item fusion, integrating semantic and collaborative information through subspace decomposition, adaptive gating, or cross-view alignment~\cite{liu2024llmesr,wang2026sage,hu2025alphafuse,wei2026faerec,hou2022unisrec,lee2026ace}.
Meanwhile, to avoid under-trained ID parameters on tail items and enable shared generalization across the catalog, (ii) another line of work drops item-specific ID embeddings and adapts item content representations directly via shared projections or geometric calibration~\cite{hou2023vqrec,li2023recformer,li2026genair}.
Generally speaking, these methods treat static item semantics and collaborative signals as two distinct views, aiming to incorporate behavioral information through either per-item ID parameters or global content transformations.
Nevertheless, the substantial discrepancy between static content similarity and progressive preference transitions renders per-item ID updates learned from sparse interactions or content projections without explicit supervision from preceding interaction contexts insufficient for aligning item representations with consecutive user needs.

We argue that adapting semantic representations with preceding interaction contexts should account for uneven behavioral evidence across items. Under long-tailed interaction distributions, frequent items accumulate sufficient preceding records to estimate reliable behavioral targets, whereas long-tail items have scarce or even zero interactions, making targets estimated directly from their own sparse contexts noisy or unavailable. Treating all estimated behavioral targets as equally reliable can propagate noise into the adaptation process. A key challenge is therefore how to learn a shared adaptation mechanism from reliable behavioral evidence that can generalize collaborative knowledge to long-tail items for both user history modeling and candidate ranking.

To address this challenge, we propose \textbf{PACERec}, a \textbf{P}rogressive \textbf{A}daptation framework that uses \textbf{C}ollaborative \textbf{E}vidence from frequent items to improve sequential \textbf{Rec}ommendation without relying on item-specific ID embeddings.
PACERec first applies semantic encoding to project frozen LLM item embeddings into initial semantic representations. For each item, it summarizes the preceding items users typically interact with before selecting it into a behavioral target. Moving an item's initial representation toward this target injects sequential transition patterns into the representation, aligning the item with the historical contexts under which users are likely to select it next. To perform this adaptation, PACERec uses a progressive adaptation flow via flow matching to learn a shared velocity field that takes item semantics as input and predicts step-by-step update directions toward the behavioral targets. To handle uneven behavioral evidence, only items with sufficient interactions are used to supervise this flow with frequency-based weights, while representation changes are constrained to prevent semantic drift. Because the learned flow is shared across the catalog and conditioned on item semantics, it can directly take the semantic representations of long-tail and cold-start items as input to infer their adapted representations, even when they lack reliable behavioral targets of their own.
For user modeling and candidate ranking, PACERec combines a sequence backbone with a recency-weighted intent residual branch over the adapted representations for preference scoring, which is fused with parameter-free semantic similarity scoring in the frozen LLM space for full-catalog ranking. Across three real-world datasets and three sequential backbones, PACERec consistently outperforms competitive baselines in both overall and tail-item recommendation. Our main contributions are summarized as follows:
\begin{itemize}[
    leftmargin=*,
    labelsep=0.5em,
    itemsep=0pt,
    topsep=2pt,
    parsep=0pt,
    partopsep=0pt
]
    \item We present a perspective on LLM-based sequential recommendation that adapts item semantics with preceding interaction contexts under uneven behavioral evidence, supporting both history modeling and candidate ranking.

    \item We propose \textbf{PACERec}, which learns a shared semantics-con\-ditioned velocity field from frequent items to progressively adapt static item semantics with preceding interaction contexts, enabling ID-free generalization to long-tail items.

    \item We conduct extensive experiments on three real-world datasets across three sequential recommendation backbones, demonstrating the superiority of PACERec in both overall and tail-item recommendation as well as the effectiveness of its key designs.
\end{itemize}

\section{Related Work}

\subsection{Sequential and Long-Tail Recommendation}
\label{sec:rw_longtail}
Sequential recommenders adopt recurrent (GRU4Rec~\cite{hidasi2016session}), convolutional (Caser~\cite{tang2018caser}), attentive (NARM~\cite{li2017narm}), self-attentive (SASRec~\cite{sasrec}, BERT4Rec~\cite{bert4rec}, S3-Rec~\cite{zhou2020s3rec}, LightSANs~\cite{fan2021lightsans}), frequency-filtering (FMLP-Rec~\cite{fmlprec}), or linear-recurrent (LRURec~\cite{yue2024lrurec}) backbones, yet learning item-specific ID embeddings leaves tail items with sparse collaborative supervision~\cite{NIPS2017_dbd22ba3}.

To enhance tail items within ID-based paradigms, CITIES~\cite{jang2020cities} infers tail embeddings from surrounding contexts, TailRec~\cite{cheng2024tailrec} filters cross-sequence neighbor contexts, MELT~\cite{kim2023melt} couples mutually reinforcing user and item branches, and TADA~\cite{dang2026tada} augments sequences via co-occurrence substitution, insertion, and mixing. While PACERec's behavioral targets resemble the transition neighbor aggregation in graph recommenders (SR-GNN~\cite{wu2019srgnn}, GCE-GNN~\cite{wang2020gcegnn}), they serve purely as fixed targets for a shared semantics-conditioned velocity field, enabling even zero-interaction items to obtain behavior-adapted representations.

\subsection{LLM-Enhanced Recommendation}
One line of LLM-enhanced recommenders retains item-specific ID parameters alongside text embeddings: LLM-ESR~\cite{liu2024llmesr} couples dual-view modeling with retrieval-augmented self-distillation, SAGE-Rec~\cite{wang2026sage} aligns semantic prototypes via user-level distillation, AlphaFuse~\cite{hu2025alphafuse} learns ID embeddings in sparse semantic subspaces, and FAERec~\cite{wei2026faerec} integrates gated fusion with multi-level alignment; however, their item-specific parameters still suffer from sparse tail supervision.

Another line adapts content representations directly without item IDs~\cite{hou2023vqrec,li2023recformer}: UniSRec~\cite{hou2022unisrec} learns transferable adaptors via parametric whitening and mixture-of-experts projections, ACE~\cite{lee2026ace} calibrates anisotropy via a linear autoencoder before training, and GenAIR~\cite{li2026genair} calibrates LLM-generated audience archetypes via feature repulsion, yet these transformations lack explicit supervision from preceding interaction contexts. Although Let It Go~\cite{pembek2025letitgo} adds norm-bounded per-item deltas to frozen content embeddings, such per-item updates remain sparsely supervised on tail items. PACERec instead predicts bounded behavioral adaptations from item semantics via a shared velocity field.

\subsection{Cold-Start and Generative Recommendation}
Cold-start methods map item content to warm collaborative embeddings via input dropout (DropoutNet~\cite{NIPS2017_dbd22ba3}), mixture-of-experts projections (Heater~\cite{zhu2020heater}), contrastive alignment (CLCRec~\cite{wei2021clcrec}), warm-to-cold distillation (ALDI~\cite{huang2023aldi}), uncertainty modeling (EUMR~\cite{li2026eumr}), or diffusion (DiffCold~\cite{zhang2026diffcold}). Whereas their targets are warm ID embeddings or teacher outputs, PACERec is fully ID-free and derives behavioral targets from preceding interaction contexts. Moreover, while diffusion and flow matching~\cite{flowmatching,liu2023rectified} typically generate next-item states (DiffuRec~\cite{li2023diffurec}, FMRec~\cite{liu2025fmrec}) or interaction vectors (FlowCF~\cite{liu2025flowcf}) at inference time, PACERec uses flow matching to learn adapted item representations that are cached offline for both history encoding and candidate scoring.

\section{Preliminaries}
\subsection{Sequential Recommendation}

Let $\mathcal I$ denote the item catalog and $\mathcal U$ the set of users. Each user $u\in\mathcal U$ has an ordered interaction history $\mathcal H_u=(i_{u,1},\ldots,i_{u,N_u})$, where $N_u$ is the history length and $i_{u,n}\in\mathcal I$ is the item at position $n$. In our setting, each item $i\in\mathcal I$ is associated with textual attributes, from which a pretrained LLM produces a frozen embedding $e_i\in\R^{d_e}$, where $d_e$ denotes the embedding dimension.

Given $\mathcal H_u$, a sequential recommender assigns each candidate item $i\in\mathcal I$ a score $s_{u,i}\equiv s(u,i\mid\mathcal H_u)$ and recommends the $K$ highest-scoring items to predict the user's next interaction.

\subsection{Flow Matching}\label{sec:prelim-fm}

To progressively transform semantic source representations toward adapted item representations, we build on flow matching~\cite{flowmatching}, which learns a time-dependent velocity field that transforms a source distribution into a target distribution. We consider the linear interpolation formulation~\cite{liu2023rectified}.

Let $(z_0,z_1)\in \R^d\times\R^d$ denote a paired source and target drawn from a joint distribution $\pi$, where $d$ denotes the dimension of the item representation space.
For $t\sim\operatorname{Uniform}(0,1)$, the interpolated state is
\begin{equation}
z_t=(1-t)z_0+tz_1.
\end{equation}
We use a source-conditioned velocity field $v_\theta(z,z_0,t)$, which additionally takes the source point $z_0$ as a condition and predicts the target velocity along the interpolation path.
The corresponding flow matching objective is
\begin{equation}
\mathcal L_{\mathrm{FM}}(\theta)
=\mathbb E_{(z_0,z_1)\sim\pi,\,t}
\left[\left\|v_\theta(z_t,z_0,t)-(z_1-z_0)\right\|_2^2\right].
\label{eq:preLoss}
\end{equation}
Because Eq.~\eqref{eq:preLoss} evaluates $v_\theta$ only at sampled interpolation states, computing this training objective requires no ODE integration.
To transform a source point, we numerically integrate the ODE $dz(t)/dt=v_\theta(z(t),z_0,t)$ with $z(0)=z_0$ from $t=0$ to $t=1$~\cite{chen2018neuralode}.
In PACERec, we instantiate this formulation by pairing each item's semantic source representation $z_0=x_i$ with a behavior-derived target $z_1=p_i$ to supervise the shared velocity field $v_\theta$, and integrate $v_\theta$ from $x_i$ to obtain its adapted representation $r_i\in\R^d$ (Section~\ref{sec:flow}).

\section{Method}

\begin{figure*}[t]
\centering
\includegraphics[width=0.92\textwidth]{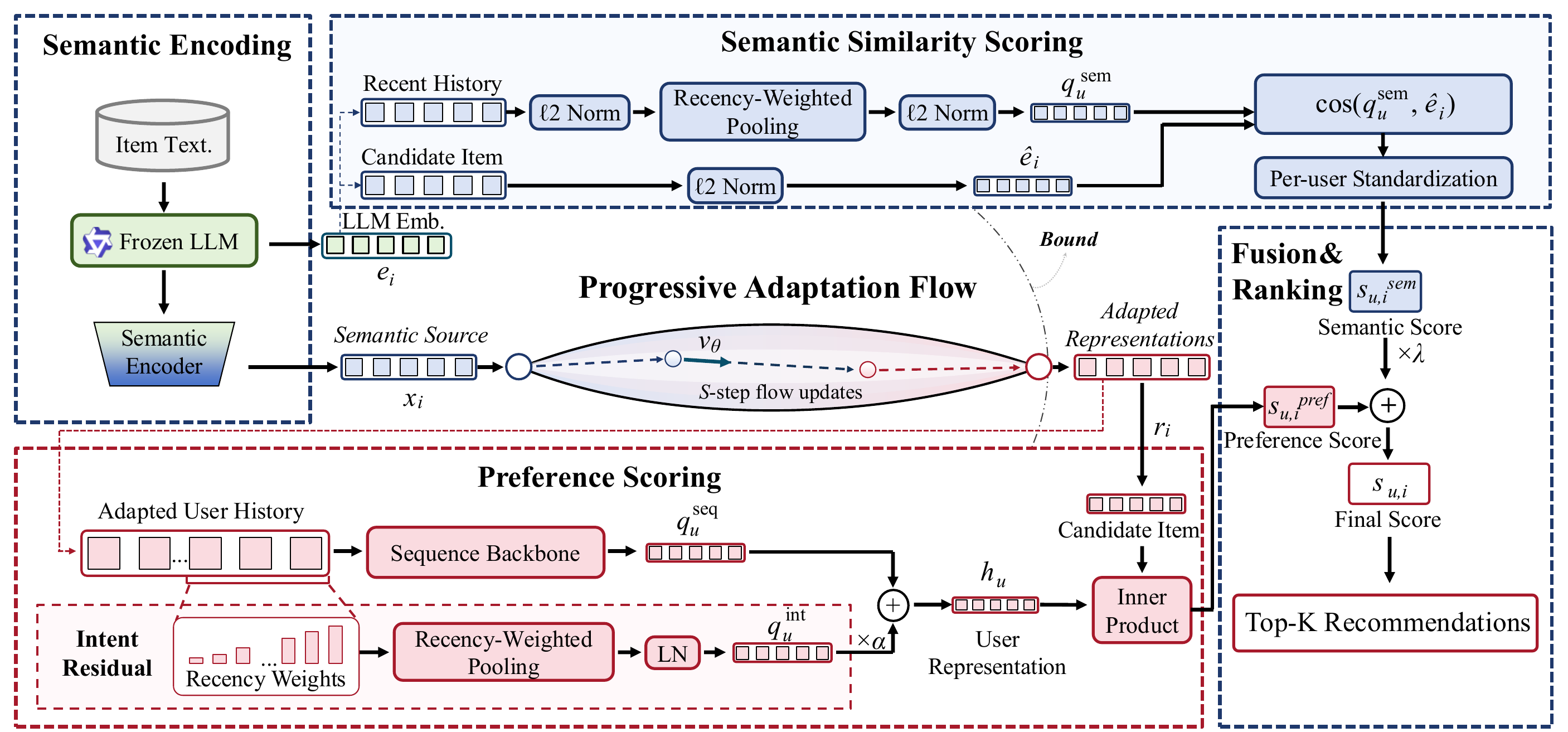}
\caption{Overview of PACERec at inference. The adaptation flow transforms each semantic source $x_i$ into an adapted representation $r_i$ for preference scoring, which is fused with semantic similarity for top-$K$ ranking.}
\Description{Architecture diagram of PACERec at inference showing Semantic Encoding, Progressive Adaptation Flow, Preference Scoring with Sequence Backbone and Intent Residual, Semantic Similarity Scoring, and Fusion and Ranking.}
\label{fig:overview}
\end{figure*}

\pace{} progressively adapts LLM-derived item representations toward collaborative behavior through two main modules, namely item representation adaptation and user representation and candidate scoring (Figure~\ref{fig:overview}).
First, for item representation adaptation (Section~\ref{sec:flow} and Figure~\ref{fig:flowdetail}), \textbf{Semantic Encoding} projects frozen LLM item embeddings into semantic source representations, and \textbf{Progressive Adaptation Flow} transforms them into adapted representations across the catalog under the supervision of preceding interaction contexts.
Second, for user representation and candidate scoring (Section~\ref{sec:scoring}), \textbf{Preference Scoring} encodes user histories over the adapted representations by combining a sequence backbone with a recency-weighted \textbf{Intent Residual}, while \textbf{Semantic Similarity Scoring} evaluates candidate similarity in the frozen LLM space for final \textbf{Fusion and Ranking}.
Finally, Section~\ref{sec:objective} presents the joint optimization objective.

\subsection{Item Representation Adaptation}\label{sec:flow}

\paragraph{\textbf{Semantic sources.}}\label{sec:source}
As shown in Figure~\ref{fig:overview}, a trainable semantic encoder maps each frozen LLM item embedding into the representation space of the recommendation backbone, giving the semantic source representation $x_i=E_\phi(e_i)\in\R^d$.
$E_\phi$ is a two-layer MLP with layer normalization at its input and output. Since $E_\phi$ relies solely on frozen embeddings derived from static item attributes, cold-start items absent from the training sequences also receive valid source representations.

\begin{figure}[t]
\centering
\includegraphics[width=\columnwidth]{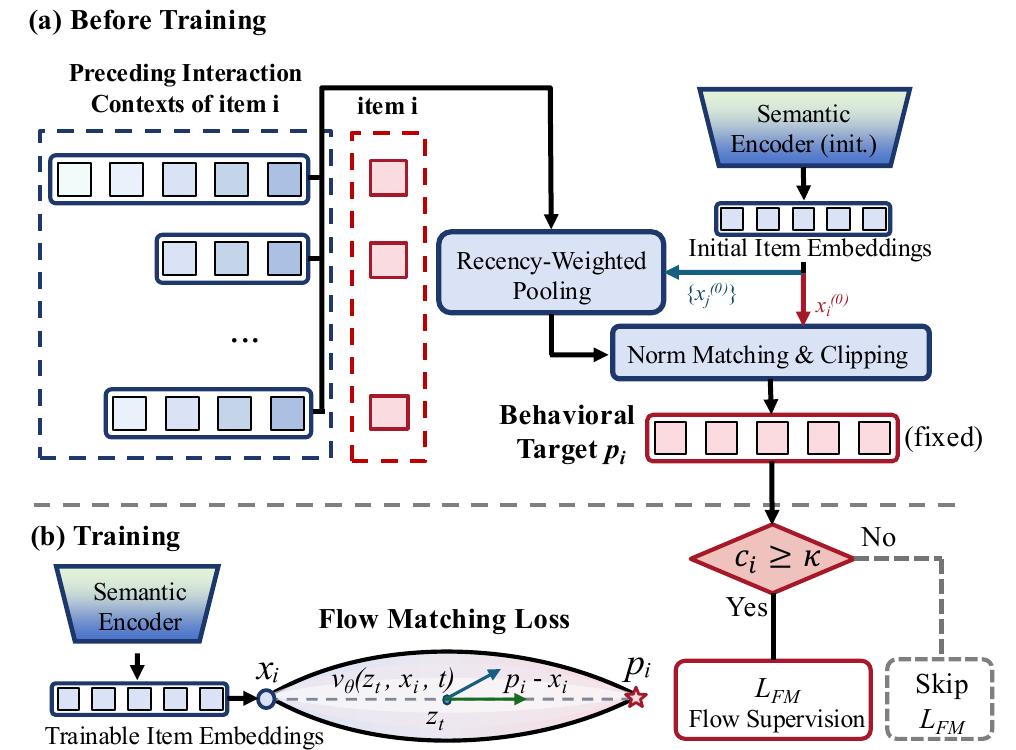}
\caption{Training of the adaptation flow: (a) offline construction of fixed behavioral targets $p_i$ before training, and (b) flow matching supervision for items with $c_i\ge\kappa$ during training.}
\Description{Two-part diagram showing (a) offline construction of fixed behavioral targets from preceding interaction contexts and (b) online confidence-weighted flow matching supervision for items meeting the evidence threshold.}
\label{fig:flowdetail}
\end{figure}

To preserve the principal semantic geometry of the frozen embeddings upon dimensionality reduction, we initialize the output projection of $E_\phi$ by least squares so that the initial sources $x_i^{(0)}$ approximate the standardized leading-$d$ PCA coordinates of $\{e_i\}_{i\in\mathcal I}$. Both the behavioral targets and the semantic anchor use these initial source representations $x_i^{(0)}$ (Figure~\ref{fig:flowdetail}(a)) as a fixed reference. During training, the encoder $E_\phi$ is optimized jointly with the recommender, so the current sources $x_i$ evolve while $e_i$ and $x_i^{(0)}$ remain fixed.

\paragraph{\textbf{Flow updates.}}

Following the source-conditioned formulation in Eq.~\eqref{eq:preLoss}, we instantiate the velocity field as $v_\theta(z,x_i,t)$, where $x_i$ denotes the semantic source and $z$ the evolving item representation.
Starting from $x_i$, we generate adapted representations through $S$ deterministic Euler updates with uniform step size $1/S$ and midpoint time inputs:
\begin{equation}
z_i^{(k+1)}=z_i^{(k)}+\tfrac{1}{S}\,v_\theta\big(z_i^{(k)},x_i,\tfrac{k+1/2}{S}\big),
\qquad z_i^{(0)}=x_i,
\label{eq:integration}
\end{equation}
for $k=0,\ldots,S-1$, with the source condition $x_i$ fixed throughout integration. We apply a smooth radial bound to the accumulated displacement $\Delta_i=z_i^{(S)}-x_i$:
\begin{equation}
r_i=x_i+\rho\tanh\!\Big(\frac{\|\Delta_i\|_2}{\rho}\Big)
\frac{\Delta_i}{\|\Delta_i\|_2},
\label{eq:endpoint}
\end{equation}
where $r_i=x_i$ when $\Delta_i=0$. This displacement bound preserves the displacement direction and keeps its magnitude below the bound radius $\rho$.
We parameterize $v_\theta$ with a sinusoidal time embedding and two residual MLP blocks, and zero-initialize its output layer so that the initial flow implements the identity mapping $r_i=x_i$.

\paragraph{\textbf{Behavioral targets.}}\label{sec:proto}
The behavioral target of item $i$ summarizes the preceding items that users interact with shortly before selecting $i$ in the training sequences, capturing the historical transition contexts under which item $i$ is likely to be selected next. Let $\mathcal C_i^{(l)}$ denote the multiset of lag-$l$ predecessors of item $i$, i.e., items occurring exactly $l$ steps before occurrences of item $i$ in the training sequences. We pool their initial source representations with recency weights:
\begin{equation}
\bar p_i=
\frac{\sum_{l=1}^{L}w_l\sum_{j\in\mathcal C_i^{(l)}}x_j^{(0)}}
{\sum_{l=1}^{L}w_l|\mathcal C_i^{(l)}|},
\label{eq:context}
\end{equation}
where $L$ is the context window and the recency weights $w_l$ decay exponentially with the lag $l$. Since averaging multiple predecessors shrinks the representation norm relative to the initial source $x_i^{(0)}$, we first apply norm matching to rescale $\bar p_i$ as $\tilde p_i=\Vert{}x_i^{(0)}\Vert{}_2\bar p_i/\Vert{}\bar p_i\Vert{}_2$.
Furthermore, to keep each supervision target consistent with the displacement bound $\rho$ in Eq.~\eqref{eq:endpoint}, we clip its displacement from $x_i^{(0)}$ to obtain the behavioral target $p_i=x_i^{(0)}+\mathrm{clip}_\rho(\tilde p_i-x_i^{(0)})$, where $\mathrm{clip}_\rho(\delta)=\delta\min(1,\rho/\Vert{}\delta\Vert{}_2)$.
When item $i$ has no preceding context in the training sequences, we set $p_i=x_i^{(0)}$.
These behavioral targets $\{p_i\}_{i\in\mathcal I}$ are pre-computed once from the initial sources $\{x_i^{(0)}\}_{i\in\mathcal I}$ before training and remain fixed, keeping them independent of the trainable encoder $E_\phi$ during optimization.

\paragraph{\textbf{Flow matching supervision.}}\label{sec:flow-supervision}
Flow matching instantiates the coupling $\pi$ of Section~\ref{sec:prelim-fm} with one pair per item, $(z_0,z_1)=(x_i,p_i)$.
Let the training count $c_i$ be the number of occurrences of item $i$ in the training sequences, counting repeated interactions.
To ensure that the velocity field is supervised by reliable behavioral targets, in each mini-batch $\mathcal B$ with distinct positive target items $\mathcal P_{\mathcal B}$, we restrict flow matching supervision to the subset $\mathcal G_{\mathcal B}=\{i\in\mathcal P_{\mathcal B}:c_i\ge\kappa\}$ whose counts reach an evidence threshold $\kappa$.
For each eligible item $i\in\mathcal G_{\mathcal B}$, we sample $t_i\sim\operatorname{Uniform}(0,1)$, construct the interpolation state $z_{t_i}=(1-t_i)x_i+t_i p_i$, and train the velocity field to predict the target velocity $p_i-x_i$:
\begin{equation}
\mathcal L_{\mathrm{FM}}=\sum_{i\in\mathcal G_{\mathcal B}}\bar\omega_i\,\frac{1}{d}\big\|v_\theta(z_{t_i},x_i,t_i)-(p_i-x_i)\big\|_2^2,
\label{eq:fm}
\end{equation}
where $\kappa$ is the largest integer such that items with $c_i\ge\kappa$ account for at least $90\%$ of training interactions (i.e., $\sum_{i:c_i\ge\kappa}c_i\ge0.9\sum_{i}c_i$), $\bar\omega_i=\omega_i/\sum_{j\in\mathcal G_{\mathcal B}}\omega_j$ normalizes the confidence weights $\omega_i=\sqrt{c_i/(c_i+c_0)}$ with a smoothing constant $c_0>0$, and $\mathcal L_{\mathrm{FM}}=0$ when $\mathcal G_{\mathcal B}$ is empty.

\subsection{User Representation and Candidate Scoring}\label{sec:scoring}

\paragraph{\textbf{Preference scoring.}}
Given user history $\mathcal H_u$, the sequence backbone $B_\psi$ encodes the ordered history of adapted item representations and returns its last hidden state $q_u^{\mathrm{seq}}$ (Figure~\ref{fig:overview}). In parallel, an intent summary pools the adapted representations with recency weights and parameter-free layer normalization:
\begin{equation}
q_u^{\mathrm{int}}=
\operatorname{LN}\!\left(
\frac{\sum_{j\in\mathcal J_u}w_j^{\mathrm{int}}r_{u,-j}}
{\sum_{j\in\mathcal J_u}w_j^{\mathrm{int}}}
\right),
\label{eq:intent}
\end{equation}
where $\mathcal J_u$ indexes all items in the retained history of at most 50 recent interactions and $r_{u,-j}$ is the adapted representation of the item $j$ steps before the most recent interaction ($j=0$ for the most recent item). The weights $w_j^{\mathrm{int}}$ decay exponentially with $j$.
Unlike the behavioral targets in Eq.~\eqref{eq:context}, which pool fixed initial sources, the intent summary pools the current adapted representations.

We form the user representation $h_u=q_u^{\mathrm{seq}}+\alpha q_u^{\mathrm{int}}$, where the mixing weight $\alpha$ is a learnable scalar bounded by a scaled sigmoid. Candidate item $i$ receives the preference score $s^{\mathrm{pref}}_{u,i}=\langle h_u,r_i\rangle$.

\paragraph{\textbf{Score fusion.}}
A semantic similarity branch scores candidates directly in the frozen LLM embedding space.
We $\ell_2$-normalize each embedding as $\hat e_i=e_i/\|e_i\|_2$, pool the normalized embeddings of the most recent items with exponentially decaying weights, and $\ell_2$-normalize the pooled vector to obtain the semantic query $q_u^{\mathrm{sem}}$.
The semantic similarity score is
$s^{\mathrm{sem}}_{u,i}=\zeta_u(\cos(q_u^{\mathrm{sem}},\hat e_i))$,
where $\zeta_u$ applies z-score standardization across the full catalog for each user.
This standardization removes user-specific shifts and scales in cosine similarities, so that a single fusion weight $\lambda$ applies to all users. The final recommendation score is
\begin{equation}
s_{u,i}=s^{\mathrm{pref}}_{u,i}+\lambda s^{\mathrm{sem}}_{u,i}.
\label{eq:fusion}
\end{equation}
The branch has no trainable parameters and is added only after training, with $\lambda$ selected on validation data.

\subsection{Joint Optimization}\label{sec:objective}

All trainable components are optimized jointly in a single stage.

\paragraph{\textbf{Ranking supervision.}}
For each training event $(u,n)$, the model predicts the item $i_{u,n}$ from the preceding items $(i_{u,1},\ldots,i_{u,n-1})$, which give the user representation $h_{u,n}$.
We use a sampled softmax~\cite{pmlr-v80-blanc18a} over preference scores with a sampling-bias correction~\cite{yi2019sampling}. With $s^{+}=\langle h_{u,n},r_{i_{u,n}}\rangle$ and $s_j=\langle h_{u,n},r_j\rangle$,
\begin{equation}
\ell_{u,n}=-\log\frac{\exp(s^{+})}{\exp(s^{+})+\sum_{j\in\mathcal N_{\mathcal B}}\exp\!\big(s_j-\log(MQ_j)\big)},
\label{eq:sampled}
\end{equation}
where $\mathcal N_{\mathcal B}$ contains $M$ negatives shared by all events in mini-batch $\mathcal B$, drawn from $Q_j\propto c_j^{\beta}$ after removing the batch positives, and $\log(MQ_j)$ corrects for the sampling distribution.
To emphasize long-tail items, we assign higher weights to infrequent positive items, $w^{+}_{u,n}=\operatorname{clip}(\sqrt{c_{\mathrm{med}}/c_{i_{u,n}}},w_{\min},w_{\max})$, where $c_{\mathrm{med}}$ is the median training count of the positive items over training events and $[w_{\min},w_{\max}]$ bounds the weights.
The ranking loss $\mathcal L_{\mathrm{rank}}$ averages $\ell_{u,n}$ over the training events in $\mathcal B$, weighted by $w^{+}_{u,n}$.
The auxiliary intent loss $\mathcal L_{\mathrm{int}}$ applies the same objective with the intent summary $q^{\mathrm{int}}_{u,n}$ in place of the user representation $h_{u,n}$. Here, the confidence weighting $\omega_i$ in Eq.~\eqref{eq:fm} emphasizes reliable behavioral targets from frequent items, while the positive-item weighting $w^{+}_{u,n}$ increases the contribution of infrequent items to ranking optimization.

\paragraph{\textbf{Semantic anchoring and joint objective.}}
We penalize drift of the trainable source representations from their initialization with the semantic anchor $\mathcal L_{\mathrm{anchor}}=\frac{1}{|\mathcal I_{\mathcal B}|}\sum_{i\in\mathcal I_{\mathcal B}}\frac{1}{d}\|x_i-x_i^{(0)}\|_2^2$, averaged over the distinct items $\mathcal I_{\mathcal B}$ in the batch. The anchor complements the displacement bound in Eq.~\eqref{eq:endpoint}: the bound constrains $\|r_i-x_i\|_2$, while the anchor penalizes $\|x_i-x_i^{(0)}\|_2^2$. The training objective is:
\begin{equation}
\mathcal L=\mathcal L_{\mathrm{rank}}
+\gamma_{\mathrm{int}}\mathcal L_{\mathrm{int}}
+\gamma_{\mathrm{FM}}\mathcal L_{\mathrm{FM}}
+\gamma_{\mathrm{anchor}}\mathcal L_{\mathrm{anchor}}.
\label{eq:loss}
\end{equation}
During training, adapted item representations are recomputed for each mini-batch using the current model parameters. While flow matching (Eq.~\eqref{eq:fm}) and the semantic anchor act only on the item adaptation side ($E_\phi$ and $v_\theta$), the ranking losses update the semantic encoder $E_\phi$, sequence backbone $B_\psi$, mixing weight $\alpha$, and velocity field $v_\theta$ through the $S$ integration steps that produce $r_i$ (Eq.~\eqref{eq:integration}).
At inference, the adapted representations $\{r_i\}_{i\in\mathcal I}$ of all catalog items are computed once and cached.

\begin{table*}[t]
\centering
\caption{
Overall and tail-item recommendation performance on three datasets.
H@10 and N@10 denote Hit@10 and NDCG@10, respectively.
Base denotes the sequence backbone trained solely on interaction data, and Tail evaluates test instances whose target items fall outside the most frequent 20\% of training items.
The best result is highlighted in \textbf{bold}, and the best baseline is
\underline{underlined}.
Impr. denotes the relative improvement of PACERec over the best baseline.
}
\label{tab:main_results}
\small
\setlength{\tabcolsep}{3.5pt}
\renewcommand{\arraystretch}{1.0}

\begin{tabular*}{\textwidth}{@{\extracolsep{\fill}}ll|cccc|cccc|cccc@{}}
\toprule

& &
\multicolumn{4}{c|}{Yelp} &
\multicolumn{4}{c|}{Food} &
\multicolumn{4}{c}{Beauty} \\

\cmidrule(lr){3-6}
\cmidrule(lr){7-10}
\cmidrule(lr){11-14}

\multirow{2}{*}{Backbone}
& \multirow{2}{*}{Method}
& \multicolumn{2}{c}{Overall}
& \multicolumn{2}{c|}{Tail}
& \multicolumn{2}{c}{Overall}
& \multicolumn{2}{c|}{Tail}
& \multicolumn{2}{c}{Overall}
& \multicolumn{2}{c}{Tail} \\

& &
H@10 & N@10 & H@10 & N@10 &
H@10 & N@10 & H@10 & N@10 &
H@10 & N@10 & H@10 & N@10 \\

\midrule

\multirow{11}{*}{SASRec}

& Base
& 0.0280 & 0.0134 & 0.0056 & 0.0026
& 0.0335 & 0.0170 & 0.0093 & 0.0053
& 0.0339 & 0.0177 & 0.0146 & 0.0085 \\

& LLM-ESR
& 0.0513 & 0.0261 & 0.0027 & 0.0012
& 0.0322 & 0.0159 & 0.0030 & 0.0013
& 0.0226 & 0.0111 & 0.0005 & 0.0002 \\

& AlphaFuse
& 0.0561 & 0.0283 & 0.0087 & 0.0038
& 0.0717 & 0.0382 & 0.0361 & 0.0193
& 0.0566 & 0.0295 & 0.0305 & 0.0159 \\

& EUMR
& 0.0359 & 0.0181 & 0.0027 & 0.0014
& 0.0345 & 0.0178 & 0.0080 & 0.0046
& 0.0338 & 0.0180 & 0.0147 & 0.0089 \\

& SAGE-Rec
& 0.0514 & 0.0270 & 0.0012 & 0.0006
& 0.0300 & 0.0151 & 0.0006 & 0.0003
& 0.0207 & 0.0106 & 0.0027 & 0.0015 \\

& ACE
& \underline{0.0563} & \underline{0.0289}
& 0.0124 & 0.0059
& 0.0714 & 0.0404
& 0.0328 & 0.0190
& \underline{0.0622} & \underline{0.0370}
& 0.0294 & 0.0182 \\

& GenAIR
& 0.0471 & 0.0236 & 0.0110 & 0.0057
& 0.0535 & 0.0290 & 0.0219 & 0.0127
& 0.0332 & 0.0173 & 0.0098 & 0.0055 \\

& FAERec
& 0.0459 & 0.0237
& \underline{0.0147} & \underline{0.0078}
& \underline{0.0781} & \underline{0.0446}
& \underline{0.0727} & \underline{0.0429}
& 0.0617 & 0.0357
& \underline{0.0658} & \underline{0.0388} \\

& DiffCold
& 0.0382 & 0.0190 & 0.0054 & 0.0030
& 0.0406 & 0.0229 & 0.0152 & 0.0091
& 0.0369 & 0.0214 & 0.0170 & 0.0108 \\

& \textbf{PACERec}
& \textbf{0.0603} & \textbf{0.0310}
& \textbf{0.0270} & \textbf{0.0139}
& \textbf{0.0903} & \textbf{0.0495}
& \textbf{0.0779} & \textbf{0.0436}
& \textbf{0.0753} & \textbf{0.0428}
& \textbf{0.0744} & \textbf{0.0436} \\

& Impr.
& 7.14\% & 7.28\% & 83.10\% & 79.12\%
& 15.60\% & 10.99\% & 7.15\% & 1.61\%
& 21.15\% & 15.55\% & 13.03\% & 12.18\% \\

\midrule

\multirow{11}{*}{FMLP-Rec}

& Base
& 0.0168 & 0.0082 & 0.0023 & 0.0012
& 0.0282 & 0.0150 & 0.0066 & 0.0043
& 0.0254 & 0.0144 & 0.0091 & 0.0067 \\

& LLM-ESR
& 0.0458 & 0.0233 & 0.0010 & 0.0004
& 0.0281 & 0.0140 & 0.0051 & 0.0025
& 0.0192 & 0.0091 & 0.0010 & 0.0005 \\

& AlphaFuse
& 0.0549 & 0.0288 & 0.0100 & 0.0048
& \underline{0.0733} & 0.0388
& 0.0390 & 0.0204
& 0.0585 & 0.0304 & 0.0316 & 0.0164 \\

& EUMR
& 0.0272 & 0.0129 & 0.0056 & 0.0030
& 0.0364 & 0.0192 & 0.0126 & 0.0072
& 0.0358 & 0.0190 & 0.0176 & 0.0107 \\

& SAGE-Rec
& 0.0454 & 0.0231 & 0.0010 & 0.0004
& 0.0289 & 0.0141 & 0.0050 & 0.0025
& 0.0256 & 0.0130 & 0.0053 & 0.0028 \\

& ACE
& \underline{0.0594} & \underline{0.0306}
& 0.0102 & 0.0051
& 0.0718 & \underline{0.0409}
& 0.0266 & 0.0156
& \underline{0.0648} & \underline{0.0389}
& 0.0282 & 0.0178 \\

& GenAIR
& 0.0521 & 0.0266 & 0.0131 & 0.0061
& 0.0605 & 0.0332 & 0.0256 & 0.0152
& 0.0397 & 0.0213 & 0.0163 & 0.0091 \\

& FAERec
& 0.0227 & 0.0119
& \underline{0.0166} & \underline{0.0088}
& 0.0711 & 0.0407
& \underline{0.0674} & \underline{0.0400}
& 0.0593 & 0.0346
& \underline{0.0661} & \underline{0.0390} \\

& DiffCold
& 0.0282 & 0.0146 & 0.0035 & 0.0021
& 0.0369 & 0.0199 & 0.0114 & 0.0067
& 0.0343 & 0.0198 & 0.0153 & 0.0099 \\

& \textbf{PACERec}
& \textbf{0.0646} & \textbf{0.0339}
& \textbf{0.0270} & \textbf{0.0151}
& \textbf{0.0931} & \textbf{0.0526}
& \textbf{0.0797} & \textbf{0.0473}
& \textbf{0.0792} & \textbf{0.0452}
& \textbf{0.0774} & \textbf{0.0452} \\

& Impr.
& 8.80\% & 10.69\% & 62.57\% & 71.95\%
& 27.00\% & 28.41\% & 18.22\% & 18.19\%
& 22.22\% & 16.20\% & 17.10\% & 15.90\% \\

\midrule

\multirow{11}{*}{LRURec}

& Base
& 0.0303 & 0.0147 & 0.0037 & 0.0023
& 0.0343 & 0.0178 & 0.0098 & 0.0057
& 0.0350 & 0.0183 & 0.0140 & 0.0083 \\

& LLM-ESR
& 0.0522 & 0.0270 & 0.0012 & 0.0004
& 0.0358 & 0.0180 & 0.0012 & 0.0007
& 0.0253 & 0.0125 & 0.0029 & 0.0015 \\

& AlphaFuse
& 0.0491 & 0.0247 & 0.0087 & 0.0038
& 0.0676 & 0.0357 & 0.0374 & 0.0195
& 0.0538 & 0.0288 & 0.0303 & 0.0161 \\

& EUMR
& 0.0382 & 0.0191 & 0.0012 & 0.0006
& 0.0359 & 0.0184 & 0.0099 & 0.0057
& 0.0369 & 0.0194 & 0.0154 & 0.0094 \\

& SAGE-Rec
& 0.0518 & 0.0269 & 0.0012 & 0.0004
& 0.0343 & 0.0177 & 0.0016 & 0.0007
& 0.0248 & 0.0122 & 0.0010 & 0.0005 \\

& ACE
& \underline{0.0544} & \underline{0.0286}
& \underline{0.0153} & 0.0075
& 0.0679 & 0.0381
& 0.0287 & 0.0168
& \underline{0.0650} & \underline{0.0389}
& 0.0322 & 0.0194 \\

& GenAIR
& 0.0438 & 0.0223 & 0.0044 & 0.0018
& 0.0458 & 0.0244 & 0.0155 & 0.0084
& 0.0283 & 0.0146 & 0.0073 & 0.0044 \\

& FAERec
& 0.0454 & 0.0230
& 0.0147 & \underline{0.0076}
& \underline{0.0749} & \underline{0.0429}
& \underline{0.0650} & \underline{0.0378}
& 0.0619 & 0.0359
& \underline{0.0640} & \underline{0.0375} \\

& DiffCold
& 0.0353 & 0.0174 & 0.0033 & 0.0017
& 0.0430 & 0.0238 & 0.0154 & 0.0094
& 0.0404 & 0.0241 & 0.0177 & 0.0112 \\

& \textbf{PACERec}
& \textbf{0.0597} & \textbf{0.0312}
& \textbf{0.0249} & \textbf{0.0130}
& \textbf{0.0896} & \textbf{0.0491}
& \textbf{0.0788} & \textbf{0.0447}
& \textbf{0.0751} & \textbf{0.0424}
& \textbf{0.0714} & \textbf{0.0418} \\

& Impr.
& 9.84\% & 9.27\% & 62.26\% & 71.01\%
& 19.72\% & 14.36\% & 21.30\% & 18.29\%
& 15.54\% & 9.00\% & 11.56\% & 11.47\% \\
\bottomrule
\end{tabular*}
\end{table*}

\begin{table*}[t]
  \centering
  \caption{Ablation results on Beauty. H@10 and N@10 denote Hit@10 and NDCG@10, respectively.}
  \label{tab:ablation}
  \small
  \setlength{\tabcolsep}{3.5pt}
  \renewcommand{\arraystretch}{1.05}
  \begin{tabular*}{\textwidth}{@{\extracolsep{\fill}}ll*{12}{c}@{}}
    \toprule
    \multicolumn{2}{c}{\multirow{3}{*}{Method}}
      & \multicolumn{4}{c}{SASRec}
      & \multicolumn{4}{c}{FMLP-Rec}
      & \multicolumn{4}{c}{LRURec} \\
    \cmidrule(lr){3-6}\cmidrule(lr){7-10}\cmidrule(lr){11-14}
    & & \multicolumn{2}{c}{Overall} & \multicolumn{2}{c}{Tail}
      & \multicolumn{2}{c}{Overall} & \multicolumn{2}{c}{Tail}
      & \multicolumn{2}{c}{Overall} & \multicolumn{2}{c}{Tail} \\
    \cmidrule(lr){3-4}\cmidrule(lr){5-6}\cmidrule(lr){7-8}
    \cmidrule(lr){9-10}\cmidrule(lr){11-12}\cmidrule(lr){13-14}
    & & H@10 & N@10 & H@10 & N@10
      & H@10 & N@10 & H@10 & N@10
      & H@10 & N@10 & H@10 & N@10 \\
    \midrule

    \multirow{3}{*}{\shortstack[l]{Structure\\Variant}}
    & w/o-SItem
      & 0.0384 & 0.0215 & 0.0109 & 0.0063
      & 0.0450 & 0.0267 & 0.0195 & 0.0121
      & 0.0394 & 0.0234 & 0.0168 & 0.0107 \\
    & w/o-Intent
      & 0.0720 & 0.0405 & 0.0581 & 0.0332
      & 0.0742 & 0.0411 & 0.0621 & 0.0351
      & 0.0659 & 0.0363 & 0.0537 & 0.0302 \\
    & w/o-Flow
      & 0.0726 & 0.0407 & 0.0683 & 0.0392
      & 0.0757 & 0.0430 & 0.0719 & 0.0413
      & 0.0714 & 0.0402 & 0.0660 & 0.0380 \\
    \midrule

    \multirow{4}{*}{\shortstack[l]{Loss\\Variant}}
    & w/o-$\mathcal{L}_{\mathrm{rank}}$
      & 0.0113 & 0.0058 & 0.0103 & 0.0053
      & 0.0703 & 0.0397 & 0.0745 & \underline{0.0441}
      & 0.0661 & 0.0396 & 0.0677 & 0.0394 \\
    & w/o-$\mathcal{L}_{\mathrm{int}}$
      & 0.0739 & 0.0408 & \underline{0.0707} & \underline{0.0407}
      & 0.0786 & \underline{0.0448} & 0.0734 & 0.0432
      & 0.0736 & \underline{0.0415} & 0.0695 & \underline{0.0405} \\
    & w/o-$\mathcal{L}_{\mathrm{FM}}$
      & \underline{0.0749} & 0.0405 & 0.0681 & 0.0384
      & 0.0781 & 0.0427 & \underline{0.0772} & 0.0416
      & \underline{0.0743} & 0.0394 & \underline{0.0696} & 0.0373 \\
    & w/o-$\mathcal{L}_{\mathrm{anchor}}$
      & 0.0747 & \underline{0.0416} & 0.0691 & 0.0399
      & \underline{0.0791} & 0.0443 & 0.0741 & 0.0423
      & 0.0740 & 0.0411 & 0.0665 & 0.0392 \\
    \midrule

    \multicolumn{2}{l}{\textbf{Full-PACERec}}
      & \textbf{0.0753} & \textbf{0.0428}
      & \textbf{0.0744} & \textbf{0.0436}
      & \textbf{0.0792} & \textbf{0.0452}
      & \textbf{0.0774} & \textbf{0.0452}
      & \textbf{0.0751} & \textbf{0.0424}
      & \textbf{0.0714} & \textbf{0.0418} \\
    \bottomrule
    \end{tabular*}
\end{table*}

\section{Experiments}
\label{sec:experiments}

\subsection{Experiment Setup}
\label{sec:experiment_setup}

\paragraph{\textbf{Datasets.}}
Following the data preprocessing setup of FAERec~\cite{wei2026faerec}, we evaluate PACERec on three real-world datasets, Beauty and Food~\cite{mcauley2015image}, and Yelp~\cite{yelp_dataset}.
All three exhibit sparse interactions and long-tailed item distributions. Beauty and Food have shorter histories than Yelp, enabling overall and tail-item evaluation under varying interaction support.
For each user, we reserve the last interaction for testing
and the penultimate interaction for validation, using earlier
interactions for training.
Detailed dataset settings and preprocessing procedures are provided in Appendix~\ref{app:dataset_settings}.

\paragraph{\textbf{Backbones and Baselines.}}
We evaluate all methods across three sequential backbones, including SASRec~\cite{sasrec}, FMLP-Rec~\cite{fmlprec}, and LRURec~\cite{yue2024lrurec}, where Base refers to the unenhanced backbone trained solely on interaction data.
We group the remaining baselines by their primary mechanisms into ID--semantic integration (AlphaFuse~\cite{hu2025alphafuse} and FAERec~\cite{wei2026faerec}), semantic knowledge transfer (LLM-ESR~\cite{liu2024llmesr} and SAGE-Rec~\cite{wang2026sage}), semantic representation refinement (ACE~\cite{lee2026ace} and GenAIR~\cite{li2026genair}), and cold-start item representation learning (EUMR~\cite{li2026eumr} and DiffCold~\cite{zhang2026diffcold}).
Detailed descriptions are provided in Appendix~\ref{app:baseline_details}.

\paragraph{\textbf{Evaluation Metrics.}}
Following prior studies~\cite{hu2025alphafuse,lee2026ace,wei2026faerec}, we report Hit Rate and Normalized Discounted Cumulative Gain at $K=10$ (Hit@10 and NDCG@10) under full-catalog ranking.
Following standard long-tail evaluation protocols~\cite{kim2023melt,liu2024llmesr,wei2026faerec}, we define the most frequent 20\% of items as head items and the remainder as tail items based solely on training interaction frequencies.
Overall metrics are computed over all test instances, while Tail metrics are computed over those whose target items belong to the tail group.

\paragraph{\textbf{Implementation Details.}}
We use frozen Qwen3-Embedding-4B~\cite{zhang2025qwen3embedding} to extract 2,560-dimensional LLM item embeddings and set the representation dimension to $d=256$.
PACERec is trained for 320 epochs using AdamW~\cite{loshchilov2019adamw} with an initial learning rate of 0.0003 and a batch size of 128.
We use $S=2$ integration steps, bound radius $\rho=2$, and $M=512$ shared negatives; the evidence threshold is $\kappa=5$ on Yelp and $\kappa=2$ on Beauty and Food.
We select the checkpoint by validation Overall NDCG@10 with $\lambda=0$, and then select the fusion weight $\lambda\in\{0,0.25,0.5,0.75,1\}$ by the sum of validation Overall and Tail NDCG@10, preferring the larger $\lambda$ on ties.
Appendix~\ref{app:details} Table~\ref{tab:hparams} lists all hyperparameters.

\subsection{Main Results}
\label{sec:main_results}
Table~\ref{tab:main_results} reports overall and tail-item recommendation performance across the three datasets and three backbones.
PACERec achieves the best results on all four metrics in all nine settings.
Compared with the strongest baseline in each setting, PACERec improves Overall Hit@10 by 7.14\% to 27.00\% and Overall NDCG@10 by 7.28\% to 28.41\%, with a mean relative gain of 13.53\% in Overall NDCG@10.
On tail items, PACERec improves Tail NDCG@10 by an average of 33.31\% across the nine settings: the gains reach 79.12\%, 71.95\%, and 71.01\% on Yelp with SASRec, FMLP-Rec, and LRURec, and range from 11.47\% to 15.90\% on Beauty and from 1.61\% to 18.29\% on Food.
PACERec thus improves both overall and tail performance, with gains varying across datasets and backbones.

The strongest baselines differ between overall and tail performance.
ACE achieves the highest baseline Overall NDCG@10 in seven of the nine configurations, whereas FAERec achieves the highest baseline Tail NDCG@10 in all configurations.
On Beauty, this contrast holds across all three backbones:
ACE achieves higher Overall NDCG@10 than FAERec, while FAERec
obtains higher Tail NDCG@10 than ACE.
By transferring preceding interaction contexts from frequent to tail items through a shared adaptation flow while preserving initial semantics, PACERec surpasses both baselines across all settings.
PACERec outperforms SASRec and ACE in held-out predecessor coverage
on Beauty (Appendix~\ref{app:visualization}).

\subsection{Ablation Study}
\label{sec:ablation}

\begin{table}[t]
\centering
\caption{Performance comparison with different LLMs and
sequence models on the Beauty dataset.}
\small
\setlength{\tabcolsep}{4.5pt}
\label{tab:backbone_comparison}
\begin{tabular*}{\columnwidth}{@{\extracolsep{\fill}}llrrrr@{}}
\toprule
\multirow{2}{*}{Backbone} & \multirow{2}{*}{Method}
& \multicolumn{2}{c}{Overall}
& \multicolumn{2}{c}{Tail} \\
\cmidrule(lr){3-4}\cmidrule(lr){5-6}
& & H@10 & N@10 & H@10 & N@10 \\
\midrule
\multicolumn{6}{l}{\textbf{Qwen3-4B}} \\
\midrule
\multirow{2}{*}{SASRec}   & FAERec  & 0.0510 & 0.0296 & 0.0431 & 0.0258 \\
         & \textbf{PACERec} & \textbf{0.0668} & \textbf{0.0374}
                  & \textbf{0.0514} & \textbf{0.0297} \\
\addlinespace
\multirow{2}{*}{FMLP-Rec} & FAERec  & 0.0481 & 0.0282 & 0.0449 & 0.0276 \\
         & \textbf{PACERec} & \textbf{0.0707} & \textbf{0.0398}
                  & \textbf{0.0584} & \textbf{0.0336} \\
\addlinespace
\multirow{2}{*}{LRURec}   & FAERec  & 0.0508 & 0.0297 & 0.0425 & 0.0256 \\
         & \textbf{PACERec} & \textbf{0.0647} & \textbf{0.0359}
                  & \textbf{0.0530} & \textbf{0.0299} \\
\midrule
\multicolumn{6}{l}{\textbf{Llama-NV-Embed-Reasoning-3B}} \\
\midrule
\multirow{2}{*}{SASRec}   & FAERec  & 0.0595 & 0.0341 & 0.0590 & 0.0341 \\
         & \textbf{PACERec} & \textbf{0.0731} & \textbf{0.0406}
                  & \textbf{0.0626} & \textbf{0.0358} \\
\addlinespace
\multirow{2}{*}{FMLP-Rec} & FAERec  & 0.0575 & 0.0333 & 0.0617 & 0.0365 \\
         & \textbf{PACERec} & \textbf{0.0753} & \textbf{0.0425}
                  & \textbf{0.0687} & \textbf{0.0398} \\
\addlinespace
\multirow{2}{*}{LRURec}   & FAERec  & 0.0596 & 0.0343 & 0.0577 & 0.0334 \\
         & \textbf{PACERec} & \textbf{0.0672} & \textbf{0.0372}
                  & \textbf{0.0595} & \textbf{0.0341} \\
\bottomrule
\end{tabular*}
\end{table}

\begin{figure}[t]
    \centering
    \includegraphics[
      width=\columnwidth
    ]{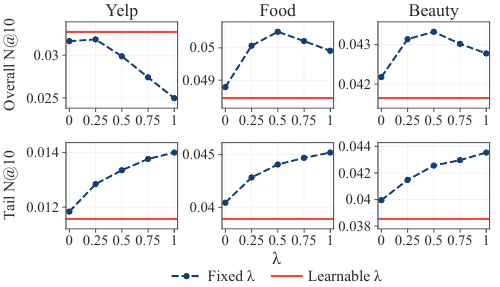}
    \caption{Effect of $\lambda$ on NDCG@10 for Yelp, Food,
    and Beauty. The top and bottom rows show overall and tail
    performance, respectively. Dashed curves represent fixed
    $\lambda$ values, while horizontal lines indicate performance
    with learnable $\lambda$.}
    \Description{Six subplots comparing fixed and learnable lambda
    on Yelp, Food, and Beauty. The top row shows overall NDCG@10,
    and the bottom row shows tail NDCG@10.}
    \label{fig:lambda-analysis}
\end{figure}

To evaluate the contribution of each key design in PACERec, we conduct ablation studies on its architectural modules, training objectives, and the semantic score fusion weight $\lambda$.
Table~\ref{tab:ablation} presents the results of the architectural and loss variants on Beauty across three backbones, where PACERec achieves the best performance across all configurations and metrics.

\paragraph{\textbf{Effect of architectural modules.}}
Among the structural variants, w/o-SItem replaces the shared adapted item representations used for history encoding and candidate scoring with trainable ID embeddings. It bypasses the adaptation flow and the semantic similarity branch, retaining semantic item representations only for the intent residual branch.
This variant reduces Tail NDCG@10 by approximately 73.2\%--85.6\%, indicating the importance of the semantic item representation pathway in the evaluated model.
Removing both the intent residual and its auxiliary loss in w/o-Intent reduces Tail NDCG@10 by approximately 22.3\%--27.8\%, supporting the effectiveness of this user modeling branch.
The w/o-Flow variant retains the semantic encoder and intent residual, directly uses the semantic source representations $x_i$ in place of the adapted representations $r_i$, and omits $\mathcal L_{\mathrm{FM}}$.
Its Tail NDCG@10 decreases by approximately 8.6\%--10.1\%, showing that behavior-guided flow adaptation provides additional benefits beyond the semantic encoder, as further illustrated by the representation visualizations in Appendix~\ref{app:visualization}.

\begin{figure*}[t]
    \centering
    \includegraphics[width=\textwidth]{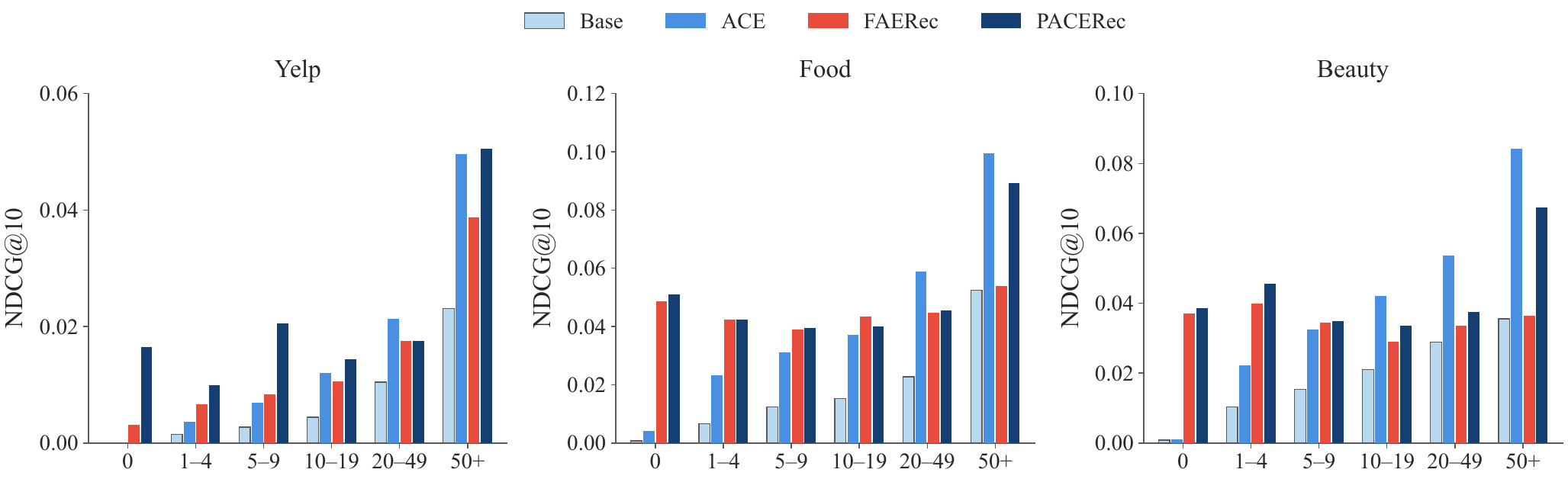}
    \caption{NDCG@10 across item frequency groups on Yelp, Food, and Beauty using SASRec.}
    \label{fig:group_analysis}
    \Description{Grouped bar charts comparing Base, ACE, FAERec, and PACERec across six item frequency groups on three datasets.}
\end{figure*}

\paragraph{\textbf{Effect of training objectives.}}
The loss variants separately remove the ranking loss ($\mathcal L_{\mathrm{rank}}$), the auxiliary intent loss ($\mathcal L_{\mathrm{int}}$), the flow matching loss ($\mathcal L_{\mathrm{FM}}$), or the semantic anchor ($\mathcal L_{\mathrm{anchor}}$). Each removal degrades performance across all three backbones. Without $\mathcal L_{\mathrm{rank}}$, the sequence backbone remains untrained while $\mathcal L_{\mathrm{int}}$ still supervises the intent summary and item representations. Without $\mathcal L_{\mathrm{FM}}$, the adaptation flow $v_\theta$ remains trained under ranking supervision alone. The resulting 8.0\%--11.9\% decrease in Tail NDCG@10 supports the value of explicit behavioral supervision.
Removing the semantic anchor also reduces Tail NDCG@10 by 6.2\%--8.5\%, highlighting the benefit of preserving initial item semantics.

\paragraph{\textbf{Effect of score fusion weight $\lambda$.}}
Figure~\ref{fig:lambda-analysis} examines the effect of the fusion weight $\lambda$, which scales the parameter-free semantic similarity score $s^{\mathrm{sem}}_{u,i}$ computed from the frozen LLM item embeddings.
As $\lambda$ increases, Tail NDCG@10 improves across all three datasets, whereas overall performance does not improve monotonically.
Yelp performs better with a small weight and shows a pronounced decline in overall performance at larger values, while Food and Beauty reach their overall peaks at moderate weights.
These results indicate that a stronger semantic scoring contribution benefits tail-item recommendation, but excessive weighting can reduce overall ranking accuracy.
Although a learnable $\lambda$ achieves higher Overall NDCG@10 on Yelp than the plotted fixed weights, it yields lower Tail NDCG@10 across all three datasets and underperforms the fixed-weight configurations on both metrics for Food and Beauty.
We therefore treat $\lambda$ as a hyperparameter selected on the validation set, jointly considering overall and tail performance to account for dataset-dependent trade-offs.

\subsection{Group Analysis}
\label{sec:group_analysis}

We further examine performance across six groups defined by item training counts $c_i$, using SASRec as the backbone.
As shown in Figure~\ref{fig:group_analysis}, PACERec performs strongly in the low-frequency groups ($0$, $1$--$4$, and $5$--$9$) and achieves the highest NDCG@10 in the zero-interaction group ($0$) across all three datasets.
These items receive no item-specific flow-matching supervision.
PACERec's advantages over Base and ACE are particularly evident for items with limited interaction support.
Compared with FAERec, PACERec also achieves higher performance in the most frequent group ($50+$) on all three datasets, showing that its benefits extend beyond sparsely observed items.
These results complement the overall and tail-item comparisons and support the effectiveness of PACERec across different levels of behavioral evidence.

\subsection{Generalizability Analysis}
The main experiments use Qwen3-Embedding-4B to extract frozen LLM item embeddings. To examine the effectiveness of PACERec across different pretrained language models, we replace the frozen LLM with Qwen3-4B~\cite{qwen2025qwen3}, a general-purpose language model, and Llama-NV-Embed-Reasoning-3B~\cite{nvidia2026nvembed}, an alternative embedding model. We evaluate both settings on the Beauty dataset with three sequence backbones. As shown in Table~\ref{tab:backbone_comparison}, PACERec consistently outperforms FAERec under both frozen LLMs, improving recommendation performance for both overall and tail items across the evaluated backbones. These results confirm the applicability of PACERec across diverse pretrained LLM item embeddings.

Beyond varying the pretrained LLMs, we further evaluate the generalizability of PACERec across additional sequence architectures and against specialized long-tail baselines in the appendix. First, Appendix~\ref{app:additional_backbones} (Table~\ref{tab:additional_backbones}) extends the evaluation to three additional backbones with distinct inductive biases, including recurrent-attentive NARM~\cite{li2017narm}, low-rank self-attentive LightSANs~\cite{fan2021lightsans}, and MLP-based gMLP~\cite{liu2021gmlp}, where PACERec yields consistent gains in both overall and tail recommendation across all three datasets. Second, Appendix~\ref{app:non_llm_baselines} (Table~\ref{tab:long-tail-comparison}) compares PACERec with four non-LLM long-tail sequential recommenders:
CITIES~\cite{jang2020cities}, MELT~\cite{kim2023melt}, TailRec~\cite{cheng2024tailrec}, and TADA~\cite{dang2026tada}. Across all three datasets and backbones, PACERec substantially surpasses these ID-based context transfer and augmentation methods, further demonstrating its broad adaptability and effectiveness.

\section{Conclusion}
\label{sec:conclusion}

In this work, we present \textbf{PACERec}, a framework that adapts frozen LLM item embeddings with collaborative evidence for sequential recommendation. A shared adaptation flow with a displacement bound learns via flow matching from fixed behavioral targets constructed from preceding interaction contexts, with count-based confidence-weighted supervision and a semantic anchor controlling representation drift. The adapted item representations support both sequence encoding and recency-weighted intent pooling for preference scoring, which is combined with parameter-free semantic similarity in the frozen LLM space for full-catalog ranking without item-specific ID embeddings. Experiments on three datasets with three sequence backbones demonstrate consistent gains in overall and tail-item recommendation, while ablations and evaluations with different frozen LLMs support the effectiveness and applicability of the framework. These results highlight the value of using preceding behavioral contexts to guide semantic adaptation under sparse and uneven interaction evidence.

\bibliographystyle{ACM-Reference-Format}
\bibliography{references}

\appendix
\setlength{\emergencystretch}{1em}

\section{Experiment Details}
\label{app:details}

\subsection{Dataset Settings}
\label{app:dataset_settings}

We conduct experiments on three real-world datasets, Beauty, Food, and Yelp.
Beauty and Food correspond to the Beauty and Grocery and Gourmet Food categories of the Amazon review dataset~\cite{mcauley2015image}, respectively.
Yelp~\cite{yelp_dataset} contains user interactions with local businesses.
Table~\ref{tab:dataset_statistics} summarizes the dataset statistics before splitting.
Sparsity denotes the proportion of unobserved user--item pairs.
The average sequence length is the mean number of interactions per user.

\begin{table}[!htbp]
    \centering
    \caption{Statistics of the three datasets.}
    \label{tab:dataset_statistics}
    \small
    \setlength{\tabcolsep}{4pt}
    \begin{tabular}{lrrrr}
        \toprule
        Dataset & \# Users & \# Items & Sparsity & Avg.\ length \\
        \midrule
        Beauty & 52,204 & 57,289 & 99.99\% & 7.56 \\
        Food   & 32,126 & 39,264 & 99.98\% & 8.57 \\
        Yelp   & 15,720 & 11,383 & 99.89\% & 12.23 \\
        \bottomrule
    \end{tabular}
\end{table}

We follow the data preprocessing setup of FAERec~\cite{wei2026faerec}.
For users with at least three interactions, we sort their interactions chronologically, reserving the second-to-last interaction for validation, the last for testing, and the remainder for training.
The maximum history length is set to 50, retaining the most recent interactions for longer histories and left-padding shorter ones with zeros.
Item frequencies are computed exclusively from training interactions (counting repeated item occurrences separately), without using validation or test data.
The most frequent 20\% of items form the head group, and the remaining items form the tail group.

\begin{table}[!htbp]
\centering
\small
\caption{PACERec hyperparameters, shared unless specified.}
\label{tab:hparams}
\begin{tabular}{@{}lp{0.65\columnwidth}@{}}
\toprule
Component & Setting \\
\midrule
Architecture
& $d=256$; hidden widths: 512/384;
2 residual MLP velocity blocks; sinusoidal time dimension: 16; dropout: 0.1 \\

Flow and targets
& $S=2$, $\rho=2$; $L=5$, target decay: 0.8 \\

Dataset-specific
& $(\kappa,c_{\mathrm{med}})=(5,47)$ for Yelp;
$(2,13)$ for Beauty and Food \\

Weighting
& $c_0=10$; positive weights: $[0.5,3]$;
$\alpha$: initial 0.25, maximum 1.5 \\

Negatives
& $M=512$ shared negatives; $\beta=0.75$ \\

History
& Intent/semantic windows: 50/10;
respective decays: 0.9/0.8 \\

Loss weights
& $(\gamma_{\mathrm{int}},\gamma_{\mathrm{FM}},
\gamma_{\mathrm{anchor}})=(0.15,0.10,0.01)$ \\

Optimizer
& AdamW; learning rate: $3\times10^{-4}$;
weight decay: $10^{-4}$; batch size: 128 \\

Schedule
& 320 epochs ($|\mathcal U|$ events/epoch);
10 warmup epochs; cosine decay to $0.1\times$;
gradient clipping: 5 \\

Validation
& Every 5 epochs;
$\lambda\in\{0,0.25,0.5,0.75,1\}$ \\
\bottomrule
\end{tabular}
\end{table}

\begin{table*}[!t]
  \centering
  \caption{Performance on additional sequential recommendation backbones. H@10 and N@10 denote Hit@10 and NDCG@10, respectively.}
  \label{tab:additional_backbones}
  \small
  \setlength{\tabcolsep}{3pt}
  \renewcommand{\arraystretch}{1.0}
  \begin{tabular*}{\textwidth}{@{\extracolsep{\fill}}c*{12}{c}@{}}
    \toprule
    & \multicolumn{4}{c}{Yelp}
    & \multicolumn{4}{c}{Food}
    & \multicolumn{4}{c}{Beauty} \\
    \cmidrule(lr){2-5}\cmidrule(lr){6-9}\cmidrule(lr){10-13}
    Backbone
    & \multicolumn{2}{c}{Overall} & \multicolumn{2}{c}{Tail}
    & \multicolumn{2}{c}{Overall} & \multicolumn{2}{c}{Tail}
    & \multicolumn{2}{c}{Overall} & \multicolumn{2}{c}{Tail} \\
    \cmidrule(lr){2-3}\cmidrule(lr){4-5}\cmidrule(lr){6-7}\cmidrule(lr){8-9}\cmidrule(lr){10-11}\cmidrule(lr){12-13}
    & H@10 & N@10 & H@10 & N@10
    & H@10 & N@10 & H@10 & N@10
    & H@10 & N@10 & H@10 & N@10 \\
    \midrule
    NARM
    & 0.0456 & 0.0231 & 0.0012 & 0.0005
    & 0.0290 & 0.0153 & 0.0035 & 0.0018
    & 0.0255 & 0.0146 & 0.0075 & 0.0040 \\
    w/ PACERec
    & \textbf{0.0618} & \textbf{0.0318} & \textbf{0.0249} & \textbf{0.0130}
    & \textbf{0.0892} & \textbf{0.0488} & \textbf{0.0759} & \textbf{0.0427}
    & \textbf{0.0729} & \textbf{0.0405} & \textbf{0.0702} & \textbf{0.0403} \\
    LightSANs
    & 0.0474 & 0.0243 & 0.0027 & 0.0014
    & 0.0458 & 0.0259 & 0.0097 & 0.0058
    & 0.0440 & 0.0271 & 0.0142 & 0.0094 \\
    w/ PACERec
    & \textbf{0.0601} & \textbf{0.0312} & \textbf{0.0266} & \textbf{0.0138}
    & \textbf{0.0918} & \textbf{0.0510} & \textbf{0.0757} & \textbf{0.0439}
    & \textbf{0.0762} & \textbf{0.0432} & \textbf{0.0731} & \textbf{0.0420} \\
    gMLP
    & 0.0369 & 0.0190 & 0.0000 & 0.0000
    & 0.0412 & 0.0241 & 0.0069 & 0.0047
    & 0.0396 & 0.0252 & 0.0086 & 0.0061 \\
    w/ PACERec
    & \textbf{0.0583} & \textbf{0.0297} & \textbf{0.0292} & \textbf{0.0149}
    & \textbf{0.0896} & \textbf{0.0494} & \textbf{0.0756} & \textbf{0.0432}
    & \textbf{0.0736} & \textbf{0.0413} & \textbf{0.0696} & \textbf{0.0400} \\
    \bottomrule
  \end{tabular*}
\end{table*}

\begin{table*}[!t]
\centering
\caption{Performance comparison with additional long-tail recommendation methods. H@10 and N@10 denote Hit@10 and NDCG@10, respectively. The best and second-best results are shown in bold and underlined, respectively.}
\label{tab:long-tail-comparison}
\small
\setlength{\tabcolsep}{3.5pt}
\renewcommand{\arraystretch}{1.0}
\begin{tabular*}{\textwidth}{@{\extracolsep{\fill}}ll*{12}{c}@{}}
\toprule
& & \multicolumn{4}{c}{Yelp}
  & \multicolumn{4}{c}{Food}
  & \multicolumn{4}{c}{Beauty} \\
\cmidrule(lr){3-6}\cmidrule(lr){7-10}\cmidrule(lr){11-14}
& & \multicolumn{2}{c}{Overall} & \multicolumn{2}{c}{Tail}
  & \multicolumn{2}{c}{Overall} & \multicolumn{2}{c}{Tail}
  & \multicolumn{2}{c}{Overall} & \multicolumn{2}{c}{Tail} \\
Backbone & Method
& H@10 & N@10 & H@10 & N@10
& H@10 & N@10 & H@10 & N@10
& H@10 & N@10 & H@10 & N@10 \\
\midrule
\multirow{5}{*}{SASRec}
& CITIES
& 0.0298 & 0.0140 & 0.0000 & 0.0000
& 0.0291 & 0.0153 & 0.0000 & 0.0000
& 0.0264 & 0.0146 & 0.0012 & 0.0005 \\
& MELT
& \underline{0.0386} & \underline{0.0191} & 0.0000 & 0.0000
& \underline{0.0450} & \underline{0.0247} & 0.0001 & 0.0001
& \underline{0.0401} & \underline{0.0236} & 0.0011 & 0.0006 \\
& TailRec
& 0.0316 & 0.0148 & \underline{0.0021} & \underline{0.0007}
& 0.0228 & 0.0120 & 0.0027 & 0.0015
& 0.0250 & 0.0143 & 0.0058 & 0.0033 \\
& TADA
& 0.0351 & 0.0178 & 0.0006 & 0.0002
& 0.0329 & 0.0186 & \underline{0.0084} & \underline{0.0055}
& 0.0303 & 0.0178 & \underline{0.0120} & \underline{0.0074} \\
& \textbf{PACERec}
& \textbf{0.0603} & \textbf{0.0310} & \textbf{0.0270} & \textbf{0.0139}
& \textbf{0.0903} & \textbf{0.0495} & \textbf{0.0779} & \textbf{0.0436}
& \textbf{0.0753} & \textbf{0.0428} & \textbf{0.0744} & \textbf{0.0436} \\
\midrule
\multirow{5}{*}{FMLP-Rec}
& CITIES
& 0.0468 & 0.0233 & 0.0004 & 0.0002
& 0.0457 & 0.0260 & 0.0068 & 0.0033
& 0.0444 & 0.0273 & \underline{0.0145} & 0.0084 \\
& MELT
& \underline{0.0597} & \underline{0.0311} & 0.0027 & 0.0012
& \underline{0.0531} & \underline{0.0301} & 0.0084 & 0.0051
& \underline{0.0490} & \underline{0.0296} & 0.0126 & 0.0074 \\
& TailRec
& 0.0464 & 0.0227 & \underline{0.0044} & 0.0021
& 0.0451 & 0.0252 & 0.0065 & 0.0039
& 0.0422 & 0.0255 & 0.0092 & 0.0056 \\
& TADA
& 0.0462 & 0.0230 & 0.0041 & \underline{0.0024}
& 0.0490 & 0.0276 & \underline{0.0108} & \underline{0.0067}
& 0.0461 & 0.0284 & 0.0143 & \underline{0.0096} \\
& \textbf{PACERec}
& \textbf{0.0646} & \textbf{0.0339} & \textbf{0.0270} & \textbf{0.0151}
& \textbf{0.0931} & \textbf{0.0526} & \textbf{0.0797} & \textbf{0.0473}
& \textbf{0.0792} & \textbf{0.0452} & \textbf{0.0774} & \textbf{0.0452} \\
\midrule
\multirow{5}{*}{LRURec}
& CITIES
& 0.0457 & 0.0230 & 0.0000 & 0.0000
& 0.0464 & 0.0271 & 0.0051 & 0.0025
& 0.0461 & 0.0288 & \underline{0.0158} & 0.0091 \\
& MELT
& \underline{0.0549} & \underline{0.0284} & 0.0004 & 0.0001
& \underline{0.0492} & \underline{0.0281} & 0.0061 & 0.0032
& 0.0461 & 0.0278 & 0.0066 & 0.0039 \\
& TailRec
& 0.0466 & 0.0232 & \underline{0.0008} & \underline{0.0003}
& 0.0478 & 0.0267 & 0.0078 & 0.0048
& 0.0432 & 0.0258 & 0.0058 & 0.0038 \\
& TADA
& 0.0450 & 0.0231 & 0.0006 & 0.0002
& 0.0473 & 0.0274 & \underline{0.0091} & \underline{0.0058}
& \underline{0.0471} & \underline{0.0289} & 0.0140 & \underline{0.0094} \\
& \textbf{PACERec}
& \textbf{0.0597} & \textbf{0.0312} & \textbf{0.0249} & \textbf{0.0130}
& \textbf{0.0896} & \textbf{0.0491} & \textbf{0.0788} & \textbf{0.0447}
& \textbf{0.0751} & \textbf{0.0424} & \textbf{0.0714} & \textbf{0.0418} \\
\bottomrule
\end{tabular*}
\end{table*}

\subsection{Baselines}
\label{app:baseline_details}

We compare PACERec with the corresponding base sequential recommendation backbones and eight representative enhancement methods.
\begin{itemize}[
    labelindent=0pt,
    leftmargin=*,
    labelsep=0.4em,
    itemsep=0pt,
    parsep=0pt,
    topsep=2pt
]

    \item AlphaFuse~\cite{hu2025alphafuse} preserves semantically rich subspaces while learning ID embeddings in semantically sparse subspaces to incorporate collaborative information.

    \item FAERec~\cite{wei2026faerec} combines adaptive gated fusion with item-level and feature-level alignment to integrate item ID and LLM-derived semantic representations.

    \item LLM-ESR~\cite{liu2024llmesr} enhances representations of long-tail users and items through semantic--collaborative dual-view modeling and retrieval-augmented self-distillation.

    \item SAGE-Rec~\cite{wang2026sage} transfers semantic knowledge into collaborative representations through item-level semantic prototype learning, user-level knowledge distillation, and global alignment.

    \item ACE~\cite{lee2026ace} uses a linear autoencoder to adjust the anisotropy of LLM embeddings while preserving their semantic structure.
    \item GenAIR~\cite{li2026genair} uses behavioral co-occurrence to calibrate item representations built from attributes and LLM-generated audience archetypes.
    \item EUMR~\cite{li2026eumr} models uncertainty in warm-item embedding statistics to simulate distribution shifts of cold-start items.
    \item DiffCold~\cite{zhang2026diffcold} trains a content-conditioned diffusion model to reconstruct warm-item ID embeddings and generate cold-start item embeddings.
\end{itemize}

\subsection{Implementation Details}
\label{app:implementation_details}

\begin{figure*}[!t]
    \centering
    \includegraphics[width=0.88\textwidth]
    {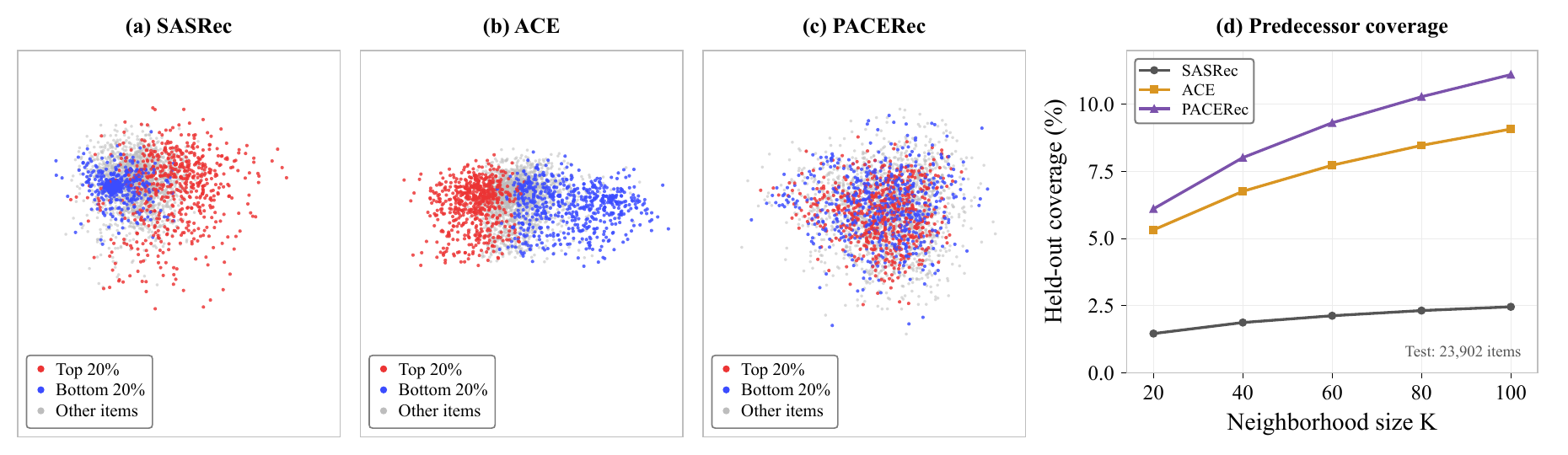}
    \caption{
        Item representation distributions and held-out predecessor
        coverage on Beauty.
        (a)--(c) PCA projections of SASRec, ACE, and PACERec,
        displaying the same 3,000 items. Red, blue, and gray denote
        the most frequent 20\%, least frequent 20\%, and remaining
        items according to training interactions, respectively.
        (d) Weighted test-predecessor coverage of cosine Top-$K$
        neighbors in the original representation spaces,
        averaged equally over 23,902 test target items.
    }
    \Description{
        Three scatter plots show item representations grouped by
        training frequency. A line chart compares predecessor
        coverage at K values from 20 to 100, with PACERec achieving
        higher coverage than ACE and SASRec.
    }
    \label{fig:beauty-item-analysis}
    \includegraphics[width=0.88\textwidth]{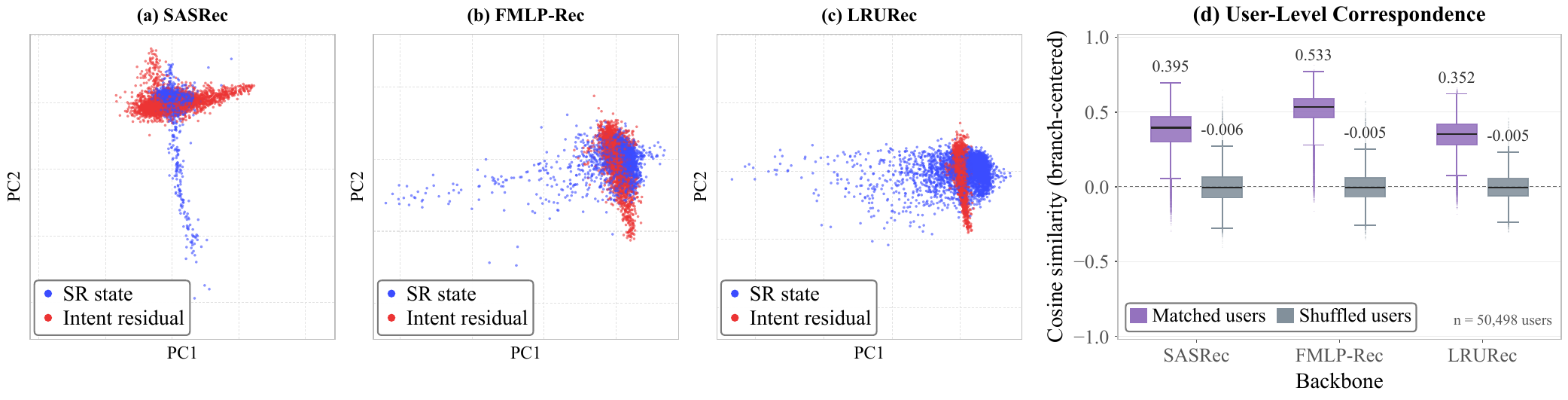}
    \caption{User representations of PACERec after removing
    each branch's mean on Beauty.
    (a--c) Joint PCA of centered SR states and intent residuals
    under SASRec, FMLP-Rec, and LRURec.
    (d) Cross-branch cosine similarity between centered representations
    for matched and randomly shuffled user pairs in the
    original feature space.}
    \Description{Four subplots on Beauty: (a-c) PCA scatter plots of centered sequence-backbone states and intent residuals for SASRec, FMLP-Rec, and LRURec; (d) boxplots of cross-branch cosine similarity for matched versus shuffled users.}
    \label{fig:user-states-centered}
\end{figure*}

All experiments are conducted on a server equipped with NVIDIA RTX A6000 GPUs.
For PACERec, checkpoint and fusion-weight selection follow Section~\ref{sec:experiment_setup}.
Table~\ref{tab:hparams} summarizes the hyperparameters, while $\kappa$ and $c_{\mathrm{med}}$ are determined from training data as described in Sections~\ref{sec:flow} and~\ref{sec:objective}.

The semantic encoder is initialized from the LLM embeddings alone.
We fit PCA on the LLM embeddings of all items and keep the leading $d$ coordinates.
With the first layer of $E_\phi$ at its random initialization, we solve a least-squares problem for the output projection and bias on 4,096 randomly sampled items, so that $E_\phi$ maps these items to their standardized PCA coordinates.
The initial sources $x_i^{(0)}$ are then computed once for all items and stored.

\section{Additional Experimental Results}
\label{app:exp}

\subsection{Additional Backbone Evaluation}
\label{app:additional_backbones}

To examine the applicability of PACERec across different sequence modeling architectures, we introduce three additional backbones, NARM, LightSANs, and gMLP.
NARM~\cite{li2017narm} combines recurrent encoding with attention to model sequential dependencies and the user's main intent within a session.
LightSANs~\cite{fan2021lightsans} employs low-rank decomposed self-attention, mapping historical items to a small set of latent interests and generating contextual representations through item-to-interest interactions.
gMLP~\cite{liu2021gmlp} uses multilayer perceptrons with spatial gating to model sequences through channel transformations and interactions across sequence positions.

\paragraph{\textbf{Overall and tail performance.}}
Table~\ref{tab:additional_backbones} shows that the improvements extend to NARM, LightSANs, and gMLP across Beauty, Food, and Yelp, beyond the backbones examined in the main experiments.
Compared with their corresponding base models, Overall NDCG@10 increases by 28.4\%--219.0\% across the nine settings, while absolute Tail NDCG@10 gains range from 0.0124 to 0.0409.
In every setting, the absolute tail gain exceeds the corresponding overall gain.
On Food and Beauty, PACERec achieves Tail NDCG@10 of 0.0400--0.0439, whereas the base models reach at most 0.0094.
On Yelp, even gMLP, whose baseline Tail NDCG@10 is reported as 0.0000, reaches 0.0149 with PACERec.
These results indicate that PACERec is effective with recurrent, attention-based, and MLP-based sequence models.

\subsection{Visualization}
\label{app:visualization}

We compare the item representation distributions and held-out predecessor coverage of SASRec, ACE, and PACERec on Beauty.
As shown in Figure~\ref{fig:beauty-item-analysis}(a)--(c), high-frequency and low-frequency items exhibit greater overlap in the two-dimensional PCA projections of PACERec.
In the original representation space (Figure~\ref{fig:beauty-item-analysis}(d)), held-out predecessor coverage is computed for each test target item and averaged uniformly, assigning equal weight to infrequent and frequent targets.
PACERec achieves higher predecessor coverage than both comparison methods across neighborhood sizes, indicating that its item neighborhoods capture more held-out predecessor relations.
These results support the design of behavioral adaptation for aligning item representations with preceding interaction contexts.

To compare the two user preference modeling paths, we remove each branch's mean and jointly apply PCA to the centered sequence-backbone states ($q_u^{\mathrm{seq}}$) and intent residuals ($\alpha q_u^{\mathrm{int}}$).
Figure~\ref{fig:user-states-centered}(a)--(c) shows partially overlapping distributions with distinct shapes across all three backbones.
In the original feature space (Figure~\ref{fig:user-states-centered}(d)), cross-branch cosine similarities are consistently higher for matched users than for randomly shuffled pairs, whose similarity distributions are centered near zero.
This indicates that the two paths retain user-level correspondence beyond global mean offsets, consistent with jointly modeling user preferences through sequence encoding and history pooling.

\subsection{Comparison with Non-LLM Baselines}
\label{app:non_llm_baselines}

We evaluate PACERec against four non-LLM long-tail sequential recommendation methods.
CITIES~\cite{jang2020cities} learns an embedding inference function from head-item interaction contexts and transfers it to tail items.
MELT~\cite{kim2023melt} improves tail-user and tail-item representations through mutually reinforcing user and item branches.
TailRec~\cite{cheng2024tailrec} enhances tail-item embeddings through lightweight contextual filtering of neighboring interactions across sequences.
TADA~\cite{dang2026tada} performs tail-aware substitution and insertion based on item co-occurrence, together with representation mixing and cross-sequence augmentation, to enrich training signals for tail users and items.

Table~\ref{tab:long-tail-comparison} shows that PACERec achieves the best overall and tail Hit@10 and NDCG@10 across all three datasets and backbones.
MELT generally delivers strong overall performance, while TADA performs well on tail items in several settings.
For example, with FMLP-Rec on Beauty, PACERec improves overall NDCG@10 by approximately 52.7\% over MELT, the strongest baseline, while tail NDCG@10 increases from 0.0096 for TADA, the strongest tail baseline, to 0.0452.

\end{document}